\documentclass[aps,prl,twocolumn,longbibliography,showpacs,amsmath,amsmath,amssymb,amsfonts,notitlepage,superscriptaddress]{revtex4-1}
\usepackage{graphicx}
\usepackage{dcolumn}
\usepackage{bm}
\usepackage{color}
\usepackage{multirow}
\usepackage[normalem]{ulem}
\usepackage{color}
\usepackage{physics}
\usepackage{bbm}
\usepackage{tikz-cd}
\usepackage{hyperref}

\begin{document}
\title{Finding the effective dynamics to make rare events typical in chaotic maps}

\author{ Ricardo Guti\'errez}
\email[]{rigutier@math.uc3m.es}
\affiliation{Grupo Interdisciplinar de Sistemas Complejos (GISC), Departamento de Matem\'aticas, Universidad Carlos III de Madrid, 28911 Legan{\'e}s, Madrid, Spain}
\author{Adri\'an Canella-Ortiz}
\affiliation{Departamento de Electromagnetismo y F\'isica de la Materia, Universidad de Granada, 18071 Granada, Spain}
\author{Carlos P\'erez-Espigares}
\email[]{carlosperez@ugr.es}
\affiliation{Departamento de Electromagnetismo y F\'isica de la Materia, Universidad de Granada, 18071 Granada, Spain}
\affiliation{Institute Carlos I for Theoretical and Computational Physics, Universidad de Granada, 18071 Granada, Spain}

\begin{abstract}
Dynamical fluctuations or rare events associated with atypical trajectories in chaotic maps due to specific initial conditions can crucially determine their fate, as the may lead to stability islands or regions in phase space otherwise displaying unusual behavior. Yet, finding such initial conditions is a daunting task precisely because of the chaotic nature of the system. In this work, we circumvent this problem by proposing a framework for finding an effective topologically-conjugate map whose typical trajectories correspond to atypical ones of the original map. This is illustrated by means of examples which focus on counterbalancing the instability of fixed points and periodic orbits, as well as on the characterization of a dynamical phase transition involving the finite-time Lyapunov exponent. The procedure parallels that of the application of the generalized Doob transform in the stochastic dynamics of Markov chains, diffusive processes and open quantum systems, which in each case results in a new process having the prescribed statistics in its stationary state. This work thus brings chaotic maps into the growing family of systems whose rare fluctuations ---sustaining prescribed statistics of dynamical observables--- can be characterized and controlled by means of a large-deviation formalism.
\end{abstract}


\maketitle

\noindent {\it Introduction---} The study of dynamical large deviations deals with fluctuations of time-averaged observables whose probabilities are exponentially suppressed in time \cite{garrahan07a, touchette09,jack20b}. This field has been enriched in recent years by the possibility of constructing effective processes where those rare fluctuations are made typical, i.e.\!  are transformed into high-probability events. This allows for controlling on demand the statistics of trajectory observables, which is especially relevant in the context of dynamical phase transitions, allowing, e.g., for the selection of certain dynamical phases that are otherwise extremely unlikely to be observed \cite{hurtadogutierrez20,hurtadogutierrez23}.  The methodology combines biased ensembles of time-averaged observables \cite{lecomte07c, garrahan09a} with the generalized Doob transform \cite{simon2009,popkov10a,jack2010,chetrite13a,chetrite15b}, and has been recently applied in stochastic systems, including lattice gas models \cite{hurtadogutierrez20,gutierrez21a,gutierrez21b,marcantoni20a}, continuum diffusive systems \cite{chetrite13a,chetrite15b,angeletti16a,nyawo18a}, and many-body systems, both classical \cite{jack2015} and quantum \cite{garrahan10a,carollo18b,marcantoni21a}.

\begin{figure}
\includegraphics[width=1.01\linewidth]{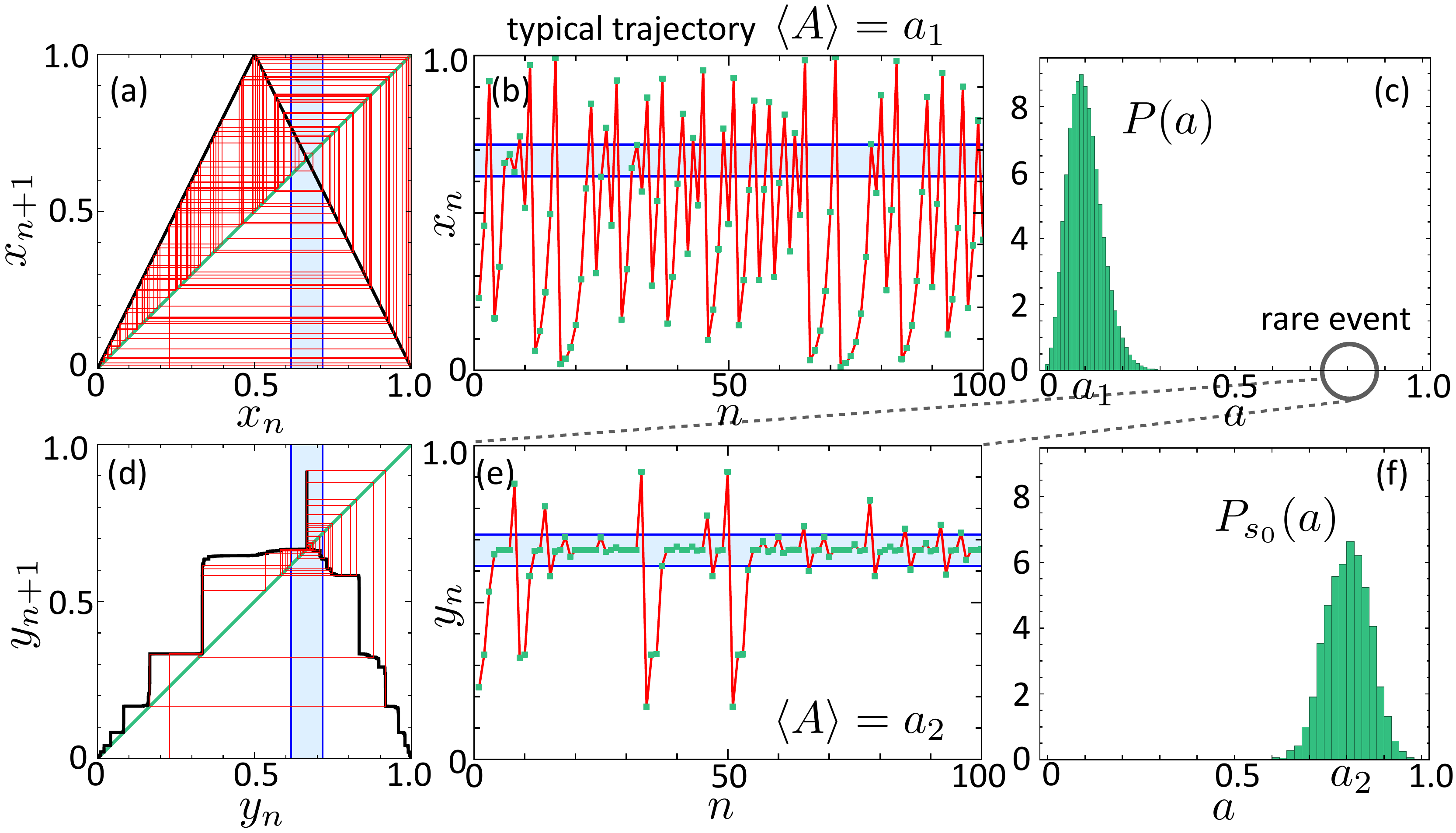}
\caption{
{\bf Rare trajectories due to the repulsive effect of an unstable fixed point are made typical.} Fluctuations of the time-averaged indicator function, $A=N^{-1}\sum_{n=0}^{N-1} \mathbb{I}_{[x^*\pm0.05]}(x_n)$, of the tent map around the unstable fixed point $x^*=2/3$. (a) Cobweb plot for $N=100$ iterations. The support of the indicator function is highlighted in light blue. (b) Trajectory illustrated in (a). (c) Histogram, $P(A=a)$, based on $10^5$ trajectories, with mean $\langle A \rangle=a_1=0.1$. (d) Cobweb plot for $N=100$ iterations of the Doob effective map with $s_0=-1$, making typical the rare fluctuation highlighted in (c). (e) Trajectory illustrated in (d). (f) Histogram, $P_{s_0}(A=a)$, based on $10^5$ trajectories of the map in (d), with mean $\langle A \rangle=a_2\approx 0.78$.}
\label{fig1}
\end{figure}

Deterministic dynamical systems are of a different nature, yet they also require a probabilistic description when their evolution is considered from a distribution of initial conditions, which is particularly relevant in the study of chaotic systems \cite{eckmann85}. In that respect, the focus of the literature on large deviations of chaotic systems from the last decades of the past century revolves around observables arising in the context of information theory and fractal geometry \cite{beck93}. A large-deviation approach to chaotic systems based on observables as general as those considered in stochastic systems, however, seems to have become available only relatively recently. Among those contributions, we highlight the Lyapunov weighted dynamics \cite{tailleur07,laffargue13a,anteneodo17a}, a computational adaptation of the cloning algorithm \cite{giardina06a,giardina11a} to Hamiltonian systems for selecting trajectories with unusual chaoticity, and the recent extension of the large-deviation formalism to general time-averaged observables in chaotic maps \cite{smith22}. Despite these advances, the adaptation of the generalized Doob transform, whereby the dynamics creating those rare trajectories is unveiled ---thus giving a powerful handle on the analysis and control of large fluctuations---, has not yet been accomplished for chaotic maps. This is a conspicuous gap in the literature that we aim to fill with the present work.

In this Letter, we propose a framework for constructing effective maps whose natural invariant measures are tailored to the statistics of general trajectory observables of a given original map. The study of rare events of chaotic maps is thus brought to a level of development that is comparable to that found in recent studies on various types of stochastic systems \cite{hurtadogutierrez20,gutierrez21a,gutierrez21b,jack2015,carollo18b}. The goal is illustrated in Fig.~\ref{fig1}, which shows an application of our framework to the tent map \cite{beck93}, $x_{n+1}=1-|1-2 x_n|$ [displayed in Fig.~\ref{fig1}(a); see Fig.~\ref{fig1}(b) for a representative trajectory corresponding to the cobweb plot]. Rare events given by trajectories with an unusually large time spent in a narrow interval centered around the unstable fixed point $x^* = 2/3$ [see Fig.~\ref{fig1}(c)], become typical in a new effective map [see Fig.~\ref{fig1}(d)], as illustrated in the histogram [Fig.~\ref{fig1}(f)] obtained from its trajectories [a representative one is displayed in Fig.~\ref{fig1}(e)].

The structure is as follows. We first show how, by extending the generalized Doob transform to the context of Frobenius-Perron operators of chaotic maps, one can generate topologically-conjugate effective maps where rare fluctuations of the original dynamics become typical.  Then we illustrate our framework by applying it to mitigate the repulsive effect of unstable periodic orbits. Finally,  we employ it to characterize dynamical phases involved  in a dynamical phase transition associated with the finite-time Lyapunov exponent in the logistic map. Concluding remarks and ideas for future work are presented at the end.

\noindent {\it Large-deviation formalism---} We consider a chaotic discrete-time dynamical system $x_{n+1} = f(x_n)$, where $f\!:\!I \to I$ is a smooth map and $I$ is some compact interval of the real line. Starting from a probability density of initial values $\alpha_0(x)$, the evolution $\alpha_{n+1}(x) = L[\alpha_n(x)]$ for $n=0,1,2,\ldots$ is given by the Frobenius-Perron operator
$L[\alpha(x)] = \int_I \alpha(y) \delta(x-f(y)) dy$, where $\delta(x)$ is a Dirac delta \cite{beck93}.
We assume that the map $f$ is ergodic with respect to an invariant measure $\rho(x) = L[\rho(x)]$. The adjoint Frobenius-Perron operator $L^\dagger$ is defined by the equality $\langle\beta, L[\alpha]\rangle = \langle L^\dagger[\beta],\alpha\rangle$, where the angular brackets denote the standard inner product, yielding $L^\dagger[\alpha(x)]  = \alpha(f(x))$; see the Supplemental Material (SM) for details \cite{SM2}. Taking $\beta(x)=\mathbbm{1}(x)=1$ above, it is clear that probability conservation, i.e. $\int L[\alpha(x)] dx=\int \alpha(x)dx=1$, implies that $L^\dagger[\mathbbm{1}(x)]=1$.

Under quite general conditions, the probability density of the time-averaged observable $A = N^{-1}\sum_{n=0}^{N-1} g(x_n)$ acquires the asymptotic large-deviation form $P(A=a) \sim e^{-N I(a)}$ for long times $N\gg 1$ \cite{oono79a,grassberger88a}. This probability concentrates around its average value, $\langle A \rangle=\int g(x)\rho(x)dx$, at a rate given by $I(a)$ ---the so-called rate function---, which is nonnegative and has a single zero located at $\langle A \rangle$ \cite{touchette09}. Thus fluctuations different from $\langle A \rangle$ become exponentially unlikely in time,  and the expansion up to second order of $I(a)$ around the mean displays Gaussian fluctuations with variance $\sigma^2=[N I''(\langle A \rangle)]^{-1}$. This is illustrated in  Fig.~\ref{fig1} (c), where the probability of the time-averaged indicator function $A=N^{-1}\sum_{n=0}^{N-1} \mathbb{I}_{[x^*\pm0.05]}(x_n)$, with $\mathbb{I}_{\Omega}(x) = 1$ if $x\in \Omega$ and zero otherwise, concentrates around $\langle A \rangle=a_1$.

The conventional method for biasing these probabilities towards specific values of $A$ is to introduce an ensemble of trajectories ---known as the $s$-ensemble \cite{garrahan07a}--- such that $P_s(a) = e^{-s N a} P(a)/Z(s)$ with $Z(s)=\int e^{-s N a} P(a)\, da$. Here $s$ is a biasing field which favors (for $s<0$) or suppresses (for $s>0$) the probability of having values larger than $\langle A \rangle$. Thus in Fig.~\ref{fig1} a suitable choice of $s=s_0=-1$ transforms the probability $P(a)$ with average $a_1 =0.1$ [Fig.~\ref{fig1} (c)], into the probability $P_{s_0}(a)$ with average $a_2 \approx 0.78$ [Fig.~\ref{fig1} (f)], which is an unusually large value in the case of the tent map. Indeed, $P(a_2) \sim e^{-N I(a_2)}$ is on the order of $10^{-18}$ for $N=100$  [see its position far into the right tail of $P(a)$ in Fig.~\ref{fig1}(c)].

In this biased ensemble, the complete statistics of the time-averaged observable $A$ for long times is given by the scaled cumulant-generating function (SCGF) $\theta(s)=\lim_{N\to \infty}N^{-1}\log Z(s)$ \cite{garrahan09a}. The latter is related to the rate function $I(a)$ by a Legendre transform, $\theta(s)=-\min_a[I(a)+sa]$ \cite{touchette09}, highlighting the analogy with the (minus) free-energy and the entropy density in equilibrium statistical mechanics, with the biasing field $s$ playing a role akin to that of the inverse temperature \cite{garrahan09a}. Since the derivatives of the SCGF provide the cumulants of the observable $A$ in the tilted distribution $P_s(a)$, the (minus) first derivative gives the average $-\theta'(s)=\langle A \rangle_s$. Thus the value of choice for $s$ is the one matching the fluctuation $a$, such that $-\theta'(s)=a$, or equivalently $I'(a)=s$. In Fig.~\ref{fig1}, $-\theta'(s_0)=a_2$ and $I'(a_2) = s_0$, while in the absence of a bias $-\theta'(0)= a_1$ and $I'(a_1) = 0$.

The SCGF is obtained from the spectral problem $L_s[r_s(x)] = e^{\theta(s)} r_s(x)$ \cite{touchette09,smith22}, where $r_s(x)$ is the right eigenfunction associated with the eigenvalue with largest real part, which is $e^{\theta(s)}$, of the so-called tilted Frobenius-Perron operator \cite{SM2}
\begin{equation}
L_s[\alpha(x)] =\!\!\! \int_I e^{-s g(y)}\alpha(y) \delta(x-f(y))\, dy =\!\!\!\!\!\!\sum_{z\in f^{-1}(x)} \!\!\!\! \frac{\displaystyle e^{-s g(z)}\alpha(z)}{|f'(z)|}\, .
\label{Lsmain}
\end{equation}
This is analogous to the definition of tilted operator for Markov chains \cite{garrahan09a} and open quantum systems \cite{garrahan10a}, and has been recently studied for chaotic maps \cite{smith22}. On the other hand, the left eigenfunction of  \eqref{Lsmain}, $l_s(x)$, satisfies $L_s^\dagger [l_s(x)] =  e^{\theta(s)} l_s(x)$, with $L_s^\dagger$ being the tilted adjoint operator, $L_s^\dagger[\alpha(x)] = e^{-s g(x)} \alpha(f(x))$, see SM \cite{SM2}. The eigenfunctions are normalized such that $\int r_s(x)dx=\int l_s(x)r_s(x)dx=1$. 
The tilted operator \eqref{Lsmain}, however, does not represent a proper physical evolution, since it does not conserve probability, $L^{\dagger}_s[\mathbbm{1}(x)]\neq 1$. Therefore it is not obvious how to derive a map, associated with $L_s$, generating the trajectories sustaining the fluctuation $a$, though such trajectories have been computationally obtained 
through the Lyapunov weighted dynamics \cite{tailleur07}. Our contribution is to show below how to obtain the effective chaotic map [as displayed in Fig.~\ref{fig1}(d)] generating those rare trajectories with $s\neq0$ [see Fig.~\ref{fig1}(e)], which follow the biased distribution $P_s(a)$ for long times [Fig.~\ref{fig1}(f)]. Such effective dynamics obtained via the Doob transform is in general difficult to construct since one needs to solve the full large-deviation problem. Various numerical schemes, such as the cloning algorithm and transition path sampling complemented with trajectory umbrella sampling \cite{nemoto16a,ray18a,klymko18a}, variational tensor networks \cite{banuls19a,causer22a,causer23a}, or machine learning techniques \cite{Oakes_2020,yan22a}, have been recently shown to converge to the effective dynamics, but always in the context of stochastic systems.

\begin{figure}
\includegraphics[width=1\linewidth]{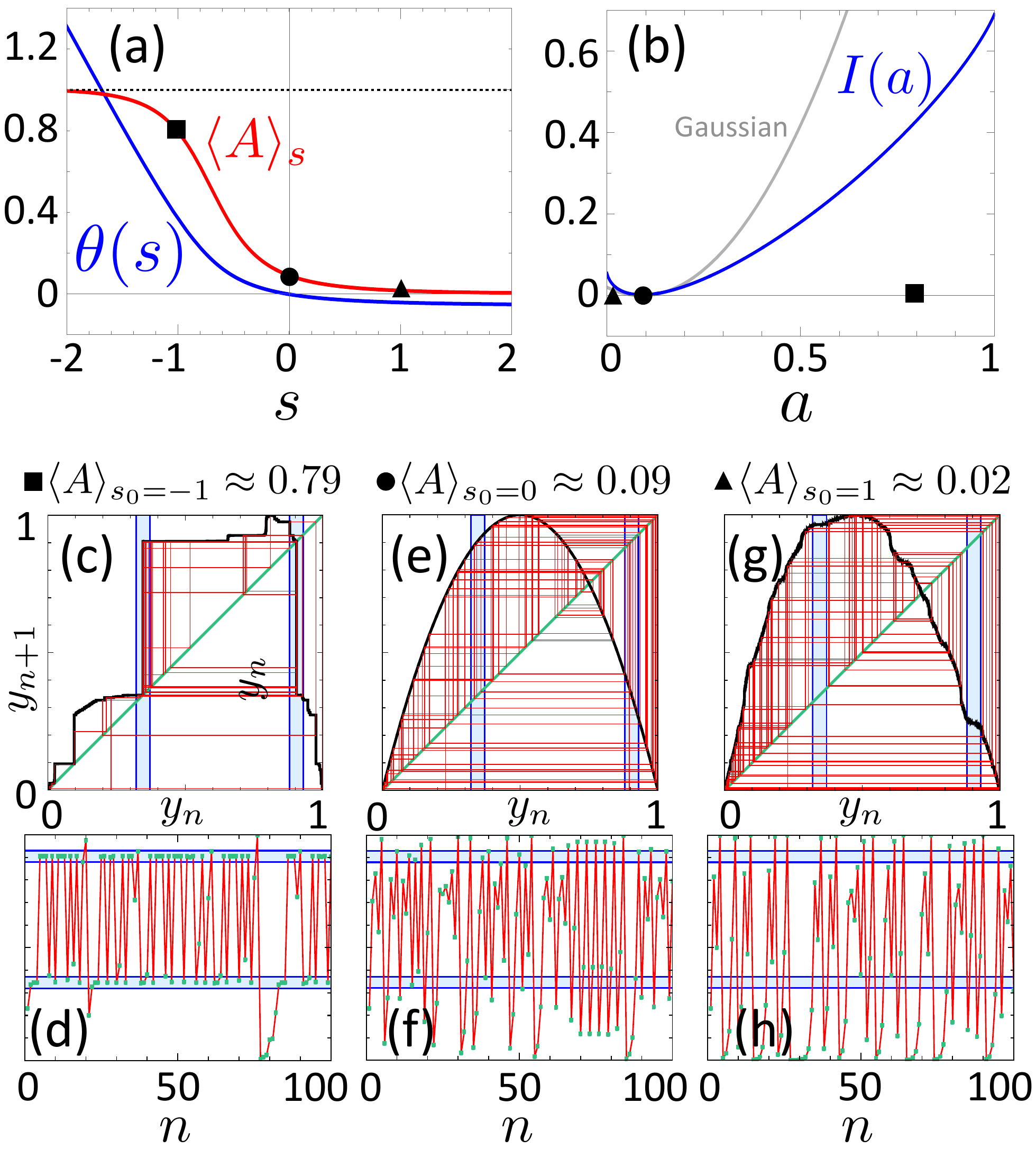}
\caption{
{\bf Rare trajectories due to the repulsive effect of unstable period-2 orbits are made typical.} Fluctuations of the time-averaged indicator function, $A=N^{-1}\sum_{n=0}^{N-1} (\mathbb{I}_{[x_-^*\pm 0.025]}(x_n)+ \mathbb{I}_{[x_+^*\pm 0.025]}(x_n))$, of the logistic map around the period-2 orbit formed by $x_{\pm}^*=(5\pm\sqrt{5})/8$.
(a) SCGF $\theta(s)$ and biased average $\langle A \rangle_s=-\theta'(s)$. The three points highlighted correspond to $s=-1$ (square), $s=0$ (circle), $s=1$ (triangle). (b) Rate function $I(a)$, and  Gaussian fluctuations around its average $\langle A \rangle$.
(c) Cobweb plot of the Doob effective map for $s_0=-1$. The support of the indicator function is highlighted in light blue. (d) Trajectory corresponding to the cobweb in (c). (e, f) Cobweb plot and trajectory of the (unbiased) logistic map ($s_0=0$). (g, h) Cobweb plot and trajectory of the Doob effective map for $s_0=1$. 
}
\label{fig2}
\end{figure}

\noindent {\it Doob operator and Doob effective map---} By analogy with the auxiliary Doob process of discrete-time stochastic systems \cite{bacco16a,coghi2019}, we define the Doob operator for a given $s=s_0$, based on the tilted operator (\ref{Lsmain}), its left eigenfunction $l_{s_0}(x)$ and the SCGF $\theta(s_0)$, as
\begin{equation}
L_{s_0}^D[\alpha(x)] = e^{-\theta(s_0)} l_{s_0}(x) L_{s_0}[\left(l_{s_0}(x)\right)^{-1}\alpha(x)].
\label{LDmain}
\end{equation}
The right eigenfunction associated with the largest eigenvalue of $L_{s_0}^D[\alpha(x)]$, which is $1$, is $\rho^D_{s_0}(x)=l_{s_0}(x) r_{s_0}(x)$, and corresponds to the stationary distribution of $L_{s_0}^D$. 
Indeed, the Doob operator (\ref{LDmain}) has the two crucial properties we sought: (i) conservation of probability, i.e. $(L^D_{s_0})^{\dagger}[\mathbbm{1}(x)]=1$, and (ii) generation of the ensemble of trajectories giving rise to the biased probability $P_{s_0}(a)$ for long times, see SM \cite{SM2} for details. 
The atypical fluctuations of the natural dynamics ($s=0$), associated with some $s_0\neq 0$ in Eq.\! (\ref{Lsmain}), thus become typical in the Doob-transformed dynamics \eqref{LDmain}.

In summary, the Doob operator (\ref{LDmain}) has a stationary state $\rho^D_{s_0}(x)$ that naturally yields the statistics for $A$  corresponding to rare fluctuations of the original dynamics, which are exponentially suppressed in $\rho(x)$, i.e.\! the invariant measure of $f$.  Yet we still need the Doob effective map, $f^D_{s_0}$, generating the atypical trajectories $y_{n+1} = f^D_{s_0}(y_n)$, which requires finding a chaotic map with a prescribed invariant measure \cite{baranovsky95}, in this case $\rho^D_{s_0}(y)$. While other maps may have the same invariant measure,  the Doob effective map $f^D_{s_0}$ is uniquely defined by the following procedure. Assuming that $\rho(x)$ and $\rho_{s_0}^D(y)$ are strictly positive and integrable (as in all the examples considered below), so that their cumulative distributions $F(x) = \int_{-\infty}^x \rho(u) du$ and $F^D_{s_0}(y) = \int_{-\infty}^y \rho_{s_0}^D(u) du$ are continuous and increasing (hence invertible) functions, the transformation that is required is $y= \gamma_{s_0}(x) = (F^D_{s_0})^{-1}(F(x))$, as it is easy to verify, see SM \cite{SM2}. Applying this transformation it is straightforward to find the Doob effective map taking into account that $y_{n+1}=f^D_{s_0}(y_{n})=f^D_{s_0}(\gamma_{s_0}(x_{n}))$ and that $y_{n+1}=\gamma_{s_0}(x_{n+1})=\gamma_{s_0}(f(x_{n}))$. From these equations we obtain $f^D_{s_0}(\gamma_{s_0}(x_{n}))=\gamma_{s_0}(f(x_{n}))$, so that the Doob effective map, which is topologically conjugate to $f$, takes the form
\begin{equation}
f^D_{s_0} = \gamma_{s_0} \circ f \circ \gamma_{s_0}^{-1}\, .
\label{effmap}
\end{equation}
The evolution is given by $f$ after a change of coordinates, $y= \gamma_{s_0}(x)$, such that $y_{n+1} = f^D_{s_0}(y_n) = \gamma_{s_0}(f(\gamma_{s_0}^{-1}(y_n)))$. 
The Doob effective map sustaining the rare event corresponding to $s_0 = -1$ in the example based on the tent map is illustrated in Fig.~\ref{fig1}(d); see the SM for the numerical method employed to obtain the eigenfunctions on which its construction is based, where it is illustrated for the doubling map, and compared with analytical and cloning-algorithm results \cite{SM2}. While $a_2$ is practically impossible to sample with the original dynamics $f$, by contrast, in the dynamics given by the effective map $f^D_{s_0}$ it is the average value. Thus the fraction of time spent in the interval $x^*\pm 0.05$ is much higher, $78\%$, as illustrated in Fig.~\ref{fig1}(e), and in the histogram of Fig.~\ref{fig1}(f).

Remarkably, while $x^* = 2/3$ is an unstable fixed point of the tent map $f$, $y^* = \gamma_{s_0}(x^*)$ (which is close to, yet different from, $2/3$) is also an unstable fixed point of the Doob map $f^D_{s_0}$. This is true in general and is imposed by the conjugacy: $f^D_{s_0}(y^*) = (\gamma_{s_0} \circ f \circ \gamma_{s_0}^{-1}) (y^*) = \gamma_{s_0}(f(x^*)) = \gamma_{s_0}(x^*) = y^*$, and $(f^D_{s_0})'(y^*) = (\gamma_{s_0} \circ f)'(x^*) (\gamma_{s_0}^{-1})'(y^*) =   \gamma_{s_0}'(x^*) f'(x^*)\left(\gamma_{s_0}'(x^*)\right)^{-1} = f'(x^*)$. Despite this, the peculiar shape of $f^D_{s_0}$ makes the trajectory spend most of the time around $x^*$ [see Fig.~\ref{fig1}(d)]. One can similarly show that a fixed point of $f^n = f\circ f \circ \cdots \circ f$ maps into a fixed point of $(f^D_{s_0})^n$ with the same stability. Those fixed points lie in periodic orbits of $f$ (with period $n$ or integers factors thereof), which is the topic we turn to next.

\noindent {\it Counterbalancing the instabilities of periodic orbits---} Unstable periodic orbits are very relevant, as many properties of chaotic systems are analyzed on such orbits embedded within chaotic attractors (see, e.g., Refs.~\cite{beck93,ott02}). Fig.~\ref{fig2} shows how to use our methodology to counterbalance the repulsive effect of unstable periodic orbits. We focus on the logistic map $f(x) = rx(1-x)$ with $r=4$ (sometimes called the Ulam map), see the black line in Fig.~\ref{fig2}(e). It has a period-2 orbit comprising $x_{\pm}^*=(5\pm\sqrt{5})/8$, which is unstable, as $(f^2)'(x_{\pm}^*) = -4$. Due to this instability, the average value of the indicator function $A=N^{-1}\sum_{n=0}^{N-1} (\mathbb{I}_{[x_-^*\pm 0.025]}(x_n)+ \mathbb{I}_{[x_+^*\pm 0.025]}(x_n))$ is only $\langle A \rangle \approx 0.09$. See Fig.~\ref{fig2}(a), which shows the SCGF $\theta(s)$, as well as its (minus) first derivative $\langle A \rangle_s$, as well as Fig.~\ref{fig2}(e) and Fig.~\ref{fig2}(f), displaying the cobweb plot and a typical trajectory of the unbiased dynamics respectively ($s=0$). As $s$ is moved towards negative (positive) values, the time average becomes larger (smaller). We will focus on $s_0=-1$, which yields $\langle A \rangle_{s_0} \approx 0.79$, associated with a much longer time spent in the vicinity of the period-2 orbit, and $s_0=1$, corresponding to $\langle A \rangle_{s_0} \approx 0.02$, for which the vicinity of the orbit is seldom visited, as displayed by Fig.~\ref{fig2}(d) and Fig.~\ref{fig2}(h), respectively. Those values of $s_0$ correspond to large deviations of $a$, well beyond the range of the Gaussian approximation, as shown in Fig.~\ref{fig2}(b).

\begin{figure}
\includegraphics[width=1.0\linewidth]{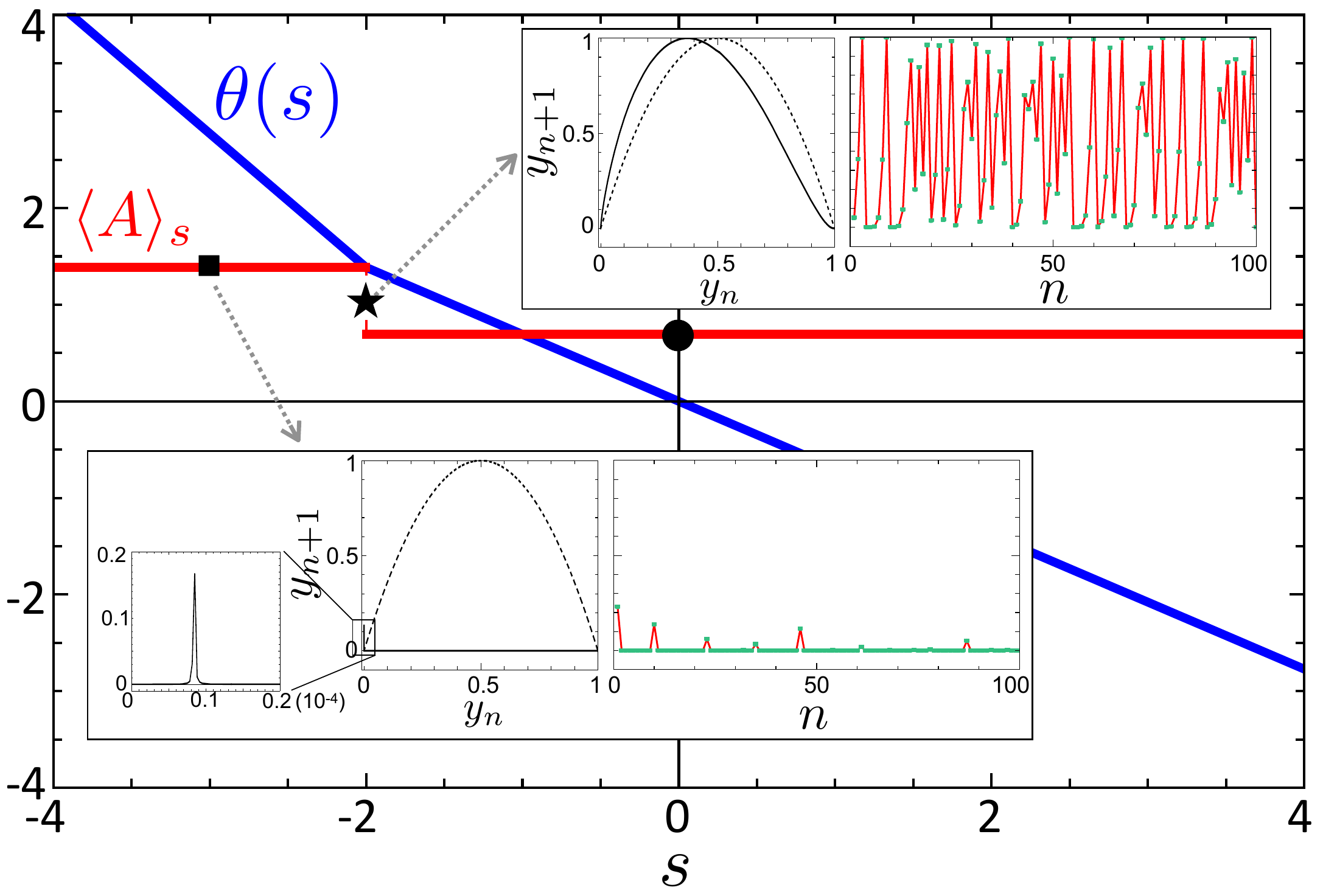}
\caption{
{\bf Characterization of phases in a DPT for the Lyapunov exponent of the logistic map.}
Main panel: SCGF $\theta(s)$ and biased average $\langle A \rangle_s=-\theta'(s)$. The three points highlighted correspond to $s=-3$ (square), $s=-2$ (star), $s=0$ (circle). The latter corresponds to the logistic map, shown in Fig.~\ref{fig2}(e) with a typical trajectory displayed in Fig.~\ref{fig2}(f). Lower inset: Doob effective map and representative trajectory for $s_0=-3$. Upper inset: Same as lower inset but at the critical point $s_0=-2$, exhibiting coexistence between both dynamical phases. In both insets the original (logistic) map is also shown (see dashed lines) }
\label{fig3}
\end{figure}

The Doob map for $s_0=-1$, see Fig.~\ref{fig2}(c), is remarkably different from the logistic map, represented in Fig.~\ref{fig2}(e). In the case of $s_0=1$ [Fig.~\ref{fig2}(g)] the difference is more subtle, yet sufficient for avoiding mapping values of $x_n$ into values of $x_{n+1}$ in the support of the indicator function. The trajectories shown in each case [Fig.~\ref{fig2}(d), (f) and (h)] correspond to the cobweb plots in the panels immediately above, and confirm all expectations.

\noindent {\it A dynamical phase transition for the Lyapunov exponent---} To conclude we focus on the timely topic of dynamical phase transitions (DPTs) \cite{garrahan09a,ates12a,jack15a,baek17a,perez-espigares18b,perez-espigares19a,banuls19a,carollo20a}. Specifically, we characterize the dynamical phases sustaining the fluctuations of the finite-time Lyapunov exponent, $A = N^{-1} \sum_{n=0}^{N-1} \ln |f'(x_n)|$, in the logistic map. For long times, the average of this fluctuating observable, which can be interpreted as a time-averaged information loss \cite{beck93}, converges to the Lyapunov exponent. The latter is $\langle A \rangle = \ln 2$, as obtained from the topological conjugacy of the logistic map and the tent map \cite{beck93,ott02}. As the tilting parameter $s$ is varied, one finds that there are just two possible values of the biased average $\langle A \rangle_s$, namely $\ln 4$ and $\ln 2$ (including obviously $s=0$). Indeed the SCGF, which for this observable is closely related to the so-called topological pressure (see e.g.~\cite{beck93}), is $\theta(s)=-2(s+1) \ln 2$ for $s \leq -2$ and $\theta(s)=- s\ln 2$ for $s \geq -2$, as discussed, with different conventions, in Refs.~\cite{tel1987,feigenbaum1989} and others therein \footnote{Specifically, $\beta F(\beta)$ in Ref.~\cite{tel1987} (which considers a topologically conjugate version of the logistic map) or $G(\beta)$ in Ref.~\cite{feigenbaum1989} correspond to minus the SCGF, $-\theta(s)$, expressed in terms of the parameter $\beta = s+1$.}. Both the SCGF $\theta(s)$ and the average $\langle A \rangle_s  = -\theta'(s)$ are displayed in Fig.~\ref{fig3}. In this case the rate function is linear, $I(a) = 2 (a -\ln 2)$, for $\ln 2 \le a \le \ln 4$, and infinite anywhere else.

We next characterize the two dynamical phases, as well as the critical point ($s=s_0 = -2$). 
For $s<-2$, the Doob effective map, presented on the left of the lower inset to Fig.~\ref{fig3} for $s_0=-3$, generates trajectories that localize in the vicinity of the point $x=0$, as displayed on the right of the same inset. There small intervals expand with a rate $\ln 4$ (instead of the common expansion rate $\ln 2$ to be found elsewhere in phase space \cite{tel1987,beck93}),
leading to $\langle A \rangle_s  = \ln 4$. On the other side of the DPT, for $s>-2$, $\langle A \rangle_s = \ln 2$, as in the unbiased dynamics ($s=0$), whose trajectories are displayed in Fig.~\ref{fig2}(f), where the region around $x\approx 0$ is hardly ever visited. Finally, the Doob effective map at the critical point $s_0 =-2$ is shown in the upper inset to Fig.~\ref{fig3}. This map generates trajectories as the one presented on the right of the inset, which exhibits a remarkable intermittency between the behavior for $s_0=-3$ and for $s_0=0$, illustrating the coexistence between dynamical phases characteristic of first-order DPTs \cite{garrahan09a,ates12a,nyawo18a,perez-espigares18b}.

\noindent {\it Concluding remarks---} 
We have developed a theoretical framework to find the effective dynamics realizing atypical trajectories in chaotic maps. Apart from its obvious interest for dynamical control purposes, it allows for the characterization of phases involved in DPTs occurring far away from the unbiased dynamics. 
While our approach has been developed for 1D systems, the formalism can be extended to cover higher-dimension maps, and perhaps also continuous-time flows. The adaptation of this framework to fluctuations at finite times by means of the finite-time Doob transform may also be feasible with currently-available techniques \cite{chetrite15b,garrahan16,carollo18b,causer22a}.

\begin{acknowledgments}
The authors thank P. Garrido, P. Hurtado, M. A. Mu\~noz and R. Hurtado-Guti\'errez for insightful discussions. The research leading to these results has been supported by Ministerio de Ciencia e Innovaci\'on (Spain), by Agencia Estatal de Investigaci\'on (AEI, Spain, 10.13039/501100011033) and by European Regional Development Fund (ERDF, A way of making Europe), through Grants PID2020-113681GB-I00, PID2021-128970OA-I00 and PID2021-123969NB-I00, by Junta de Andaluc\'ia (Spain)-Consejer\'ia de Econom\'ia y Conocimiento 2014-2020 through grant A-FQM-644-UGR20, and by Comunidad de Madrid (Spain) under the Multiannual Agreement with UC3M in the line of Excellence of University Professors (EPUC3M23), in the context of the V Plan Regional de Investigaci\'on Cient\'{\i}fica e Innovaci\'on Tecnol\'ogica (PRICIT). We are grateful for the the computing resources and related technical support provided by PROTEUS, the supercomputing center of Institute Carlos I in Granada, Spain.
\end{acknowledgments}

\onecolumngrid
\newpage

\renewcommand\thesection{S\arabic{section}}
\renewcommand\theequation{S\arabic{equation}}
\renewcommand\thefigure{S\arabic{figure}}
\setcounter{equation}{0}
\setcounter{figure}{0}

\begin{center}
{\Large{SUPPLEMENTAL MATERIAL:\\\bigskip Finding the effective dynamics to make rare events typical in chaotic maps}}
\end{center}

\maketitle

\section{Theoretical framework}

\subsection{Frobenius-Perron operator and its adjoint}

We consider a chaotic discrete-time dynamical system whose evolution $x_{n+1} = f(x_n)$ is given by a smooth map $f: I \to I$, where $I$ is some compact interval of the real line. Considering a probability distribution of initial values given by the density $\alpha_0(x)$ ($\int_I \alpha_0(x)\, dx = 1$), the time evolution of this density is given by the Frobenius-Perron operator \cite{beck93},
\begin{equation}
L[\alpha(x)] = \int_I \alpha(y) \delta(x-f(y))\, dy = \sum_{z\in f^{-1}(x)} \frac{\alpha(z)}{|f'(z)|},
\label{L}
\end{equation}
in the sense that $\alpha_{n+1}(x) = L[\alpha_n(x)]$ for $n=0,1,2,\ldots N$. In Eq.~(\ref{L}), $\delta(x)$ is a Dirac delta, and $f^{-1}(x)$ is shorthand for the set of pre-images of $x$ under the (generally non-invertible) map $f$. We assume that the map $f$ is ergodic with respect to an invariant measure
\begin{equation}
\rho(x) = L[\rho(x)] = \sum_{z\in f^{-1}(x)} \frac{\rho(z)}{|f'(z)|}\, .
\label{rho}
\end{equation}

Considering the standard inner product between real functions $\alpha(x)$ and $\beta(x)$, $\langle\alpha, \beta\rangle = \int_I \alpha(x) \beta(x)\, dx$, one can define the adjoint operator $L^\dagger$ by the relation $\langle\beta, L[\alpha]\rangle = \langle L^\dagger[\beta],\alpha\rangle$:
\begin{align}
\langle\beta, L[\alpha]\rangle &= \int_I \beta(x) L[\alpha(x)] dx = \int_I \beta(x) \left[\int_I \alpha(y) \delta(x-f(y))\, dy \right] dx\nonumber \\
&= \int_I \alpha(y) \left[\int_I \beta(x) \delta(x-f(y))\, dx \right] dy = \int_I \alpha(y) L^\dagger[\beta(y)]\, dy = \langle L^\dagger[\beta],\alpha\rangle,
\label{dag1}
\end{align}
where we have used Fubini's theorem. Thus the adjoint or dual operator is
\begin{equation}
L^\dagger[\alpha(x)]  = \int_I \alpha(y) \delta(y-f(x))\, dy = \alpha(f(x)).
\label{dag2}
\end{equation}

\subsection{Statistics of trajectory observables and tilted operators}

Under quite general conditions, the probability density of a time-averaged dynamical observable $A = N^{-1}\sum_{n=0}^{N-1} g(x_n)$ for long times acquires the large-deviation form $P(A=a) \sim e^{-N I(a)}$, where $I(a)$ is the so-called rate function. The latter is non-negative and equal to zero at the mean value, $\langle A \rangle=\int g(x) \rho(x) dx$, and has small Gaussian fluctuations around it (given by the quadratic expansion of $I(a)$ around its average), and larger (generally) non-Gaussian fluctuations corresponding to large fluctuations, i.e.\! rare events \cite{touchette09}. In order to obtain $I(a)$, it is useful to consider the so-called tilted operator, recently studied in the context of chaotic maps \cite{smith22},
\begin{equation}
L_s[\alpha(x)] = \int_I e^{-s g(y)} \alpha(y) \delta(x-f(y))\, dy 
\label{Ls}
\end{equation}
since the statistics of $A$ can be extracted from its largest eigenvalue, as explained below. Equation \eqref{Ls} can be viewed as a Frobenius-Perron operator that is biased towards atypical values of $A$ by tilting with a conjugate parameter $s$. For $s>0$, it biases the dynamics towards values smaller than $\langle A \rangle$, and for $s<0$, towards values larger than $\langle A \rangle$. However, unlike the Frobenius-Perron operator, which is obtained for $s=0$, this tilted operator (for $s\neq 0$) is not probability-conserving (as shown below). Such operators have been recently studied in \cite{smith22}, and are closely related to the `escort distributions' discussed, e.g., in Ref.~\cite{beck93}. 

Following the steps of the derivation of $L^\dagger$ given by \eqref{dag1}-\eqref{dag2}, the tilted adjoint or dual operator $L_s^\dagger$ must satisfy
\begin{align}
\langle\beta, L_s[\alpha]\rangle &= \int_I \beta(x) L_s[\alpha(x)] dx = \int_I \beta(x) \left[\int_I e^{-s g(y)} \alpha(y) \delta(x-f(y))\, dy \right] dx\nonumber \\
&= \int_I \alpha(y) \left[e^{-s g(y)}  \int_I \beta(x) \delta(x-f(y))\, dx \right] dy = \int_I \alpha(y) L_s^\dagger[\beta(y)]\, dy = \langle L_s^\dagger[\beta],\alpha\rangle.
\end{align}
Thus the tilted adjoint operator is
\begin{equation}
L_s^\dagger[\alpha(x)]  =e^{-s g(x)} \int_I \alpha(y) \delta(y-f(x))\, dy = e^{-s g(x)} \alpha(f(x)).
\end{equation}

The form of the tilted operator \eqref{Ls} can be derived by analogy with the reasoning provided for the case of jump processes in Ref.~\cite{garrahan09a}. We consider the probability density $\pi_{N}(x,a)$ of being in state $x$ after $N$ time steps, having measured a value $a$ of the observable $A=N^{-1}\sum_{n=0}^{N-1} g(x_n)$. This probability density can be related to the probability density of the previous time step as follows:
\begin{equation}
\pi_{N}(x,a) = \int_I \pi_{N-1}(y,a- N^{-1} g(y))\, \delta(x-f(y))\, dy\,.
\label{twosteps}
\end{equation}
Thus the probability of measuring a value $a$ in a trajectory of $N$ time steps is 
\begin{equation}
P(a)=\int_I \pi_{N}(x,a) \, dx\,.
\label{proba}
\end{equation}
\\
By an application of the Laplace transform 
\begin{equation}
\tilde{\pi}_{N}(x,s) = \int e^{-s N a} \pi_{N}(x,a)\, da,
\label{Ltrans}
\end{equation}
on both sides of Eq.~(\ref{twosteps}), where the integral is assumed to converge, we obtain
\begin{align}
\nonumber
\tilde{\pi}_{N}(x,s) &= \int_I \left( \int e^{-s N a} \pi_{N-1}(y,a-N^{-1}g(y))\, da\right)\, \delta(x-f(y))\, dy\\ 
&=\int_I e^{-s g(y)} \left( \int e^{-s N a'} \pi_{N-1}(y,a')\, da'\right)\, \delta(x-f(y))\, dy = \int_I e^{-s g(y)} \tilde{\pi}_{N-1}(x,s)\, \delta(x-f(y))\, dy. 
\label{twosteps-22}
\end{align}
We then see that the tilted operator \eqref{Ls} gives the time evolution of $\tilde{\pi}_{N}(x,s)$, i.e. $\tilde{\pi}_{n+1}(x,s)=L_s[\tilde{\pi}_{n}(x,s)]$. 

The statistics of the fluctuating observable $A$ can be retrieved from the moment generating function $Z(s)=\int e^{-sNa}P(a)da$, which by virtue of Eqs. \eqref{proba} and \eqref{Ltrans}, can be written as $Z(s)=\int_I \tilde{\pi}_{N}(x,s) dx$. Thus by writing $L_s[\rho(x)]$ in terms of its right and left eigenfunctions, $r_s^j(x)$ and $l_s^j(x)$, as $L_s[\rho(x)]=\sum_{j\ge 0} \lambda_j(s) r_s^j(x)(\int_I l_s^j(x)\rho(x)dx)$ with $\lambda_j(s)$  being the corresponding eigenvalues (ordered in decreasing value of their real part, $\Re[\lambda_0(s)]>\Re[\lambda_1(s)]\ge \cdots$), we obtain $Z(s)=\int_I L_s^N[\alpha_0(x)] dx=\sum_{j\ge 0} \lambda^N_j(s)  (\int_I r_s^j(x)dx)(\int_I l_s^j(x)\alpha_0(x)dx)$, when we start by an initial probability density $\alpha_0(x)$. Therefore in the long time limit $N\gg1$, $Z(s)$ follows a large deviation form $Z(s)\sim e^{N\theta(s)}(\int_I r_s(x)dx)(\int_I l_s(x)\alpha_0(x)dx)$, where $r_s(x)$ and $l_s(x)$ stand for the right and left eigenfunctions, respectively (we have removed the $0$ superscript for simplicity), associated with the eigenvalue with largest real part, $\lambda_0(s)=e^{\theta(s)}$,  
\begin{equation}
L_s[r_s(x)] = e^{\theta(s)} r_s(x)\, ,
\label{Lsrs}
\end{equation}
\begin{equation}
L_s^\dagger [l_s(x)] = e^{\theta(s)} l_s(x)\, ,
\label{Lsls}
\end{equation}
normalized such that $\int_I r_s(x)dx=1$ and $\int_I l_s(x)r_s(x)dx=1$. Then $\theta(s)=\lim_{N\to \infty} N^{-1} \ln Z(s)$ is the scaled cumulant generating function (SCGF), which is related to the rate function through a Legendre transform, $I(a)=-\min_s[\theta(s)+s a]$ \cite{touchette09}.

For $s=0$, the largest eigenvalue is $e^{\theta(0)} = 1$ ($\theta(0) = 0$): as $\alpha_{n+1}(x) = L[\alpha_{n}(x)]$ relaxes to the natural invariant measure $\rho(x)$, we have an eigenvalue of $1$ for that particular eigenfunction, $\rho(x) = L[\rho(x)]$, while others have associated eigenvalues smaller than $1$ (so the contributions of these modes to a generic initial distribution $\alpha_0(x)$ decay across time). Eq.\! (\ref{Lsrs}) reduces in that case to Eq.~(\ref{rho}), where $L_0 = L$ [see Eqs.\! (\ref{L}) and (\ref{Ls})] and $r_0(x) = \rho(x)$ is the natural invariant measure of the chaotic map $f$. On the other hand, the left eigenfunction $l_s(x)$ associated with the largest eigenvalue of $L_s$ is the corresponding right eigenfunction of the adjoint tilted operator $L^\dagger_s$, since $\langle l_s, L_s[\beta]\rangle = e^{\theta(s)}\langle l_s, \beta\rangle$  must be equal to $\langle L^\dagger_s[l_s], \beta \rangle$ for any $\beta$. 
For $s=0$, again we have $\theta(0) = 0$, so $\langle l_0, L[\beta]\rangle = \langle l_0, \beta\rangle$ for any $\beta$, which can only be satisfied for $l_0(x) = \mathbbm{1}(x) = 1$, and amounts to the conservation of probability under the action of the Frobenius-Perron operator $L$. In terms of the tilted adjoint operator, conservation of probability thus means:
\begin{equation}
L^\dagger [\mathbbm{1}(x) ] = \mathbbm{1}(f(x)) = 1.
\end{equation}

\subsection{Doob operator}

At this stage, following the analogy with the study of the auxiliary discrete-time Doob process in stochastic systems \cite{bacco16a,coghi2019}, we define the Doob operator, which, for a given value of $s=s_0$, when applied on the normalized function $\alpha(x)$, yields
\begin{align}
L_{s_0}^D[\alpha(x)] &= e^{-\theta(s_0)} l_{s_0}(x) L_{s_0}[\left(l_{s_0}(x)\right)^{-1}\alpha(x)]. 
\label{LD}
\end{align}
The right eigenfunction associated with the largest eigenvalue, which is $1$, is given by $\rho^D_{s_0}(x) = l_{s_0}(x) r_{s_0}(x)$, which corresponds to the invariant measure of $L_{s_0}^D$, since $L_{s_0}^D[l_{s_0}(x) r_{s_0}(x)] = e^{-\theta({s_0})} l_{s_0}(x) L_{s_0}[r_{s_0}(x)] = l_{s_0}(x) r_{s_0}(x)$.\\

\noindent The importance of this operator is due to the following two properties:
\begin{itemize}
\item[1)] it is a probability conserving operator (which $L_s$ is not),
\item[2)] it makes rare events typical, in the sense that the statistics associated with a rare fluctuation observed for $s_0 \neq 0$ (as encoded in the SCGF $\theta(s)$) becomes associated with the natural stationary dynamics of the Doob operator. This implies that the Doob operator generates the ensemble of trajectories giving rise to the biased probability $P_{s_0}(a)$ for long times.
\end{itemize}

Property 1) can be explicitly verified by checking that the Doob operator does indeed satisfy the requirement that $\int_I L_{s_0}^D[\beta(x)] = \int_I (L_{s_0}^D)^\dagger[\mathbbm{1}(x)] \beta(x) dx = 1$ for any normalized $\beta(x)$, hence $(L_{s_0}^D)^\dagger[\mathbbm{1}(x)] = \mathbbm{1}(x) = 1.$ The adjoint Doob operator $L_{s_0}^{D,\dagger}$ is defined by
\begin{align}
\langle\beta, L_{s_0}^D[\alpha]\rangle &= \int_I \beta(x) L_{s_0}^D[\alpha(x)] dx = \int_I \beta(x) e^{-\theta({s_0})} l_{s_0}(x) L_{s_0}[\left(l_{s_0}(x)\right)^{-1}\alpha(x)] dx\nonumber \\
&= e^{-\theta({s_0})} \int_I L_{s_0}^\dagger[\beta(x)\,  l_{s_0}(x)] \left(l_{s_0}(x)\right)^{-1}\alpha(x) dx = \langle L_{s_0}^{D,\dagger}[\beta],\alpha\rangle.
\end{align}
This yields the following expression
\begin{equation}
L_{s_0}^{D,\dagger}[\alpha(x)] =  e^{-\theta({s_0})} \left(l_{s_0}(x)\right)^{-1}  L_{s_0}^\dagger[l_{s_0}(x)\, \alpha(x)]  = e^{-[\theta({s_0})+{s_0} g(x)]} \left(l_{s_0}(x)\right)^{-1}  l_{s_0}(f(x))\, \alpha(f(x)),
\end{equation}
which indeed satisfies $(L_{s_0}^D)^\dagger[\mathbbm{1}(x)] = e^{-\theta({s_0})} \left(l_{s_0}(x)\right)^{-1}  L_{s_0}^\dagger[l_{s_0}(x)\, \mathbbm{1}(x)] = e^{-\theta({s_0})} \left(l_{s_0}(x)\right)^{-1} e^{\theta({s_0})}\, l_{s_0}(x) = 1.$ Thus the Doob operator is a probability-conserving operator,  unlike the tilted operator: $L_{ s}^\dagger[\mathbbm{1}(x)]  = e^{-{s} g(x)} \mathbbm{1}(f(x)) = e^{-{s} g(x)} \neq 1$ (for ${ s}\neq 0$).
\\
\\
\noindent Property 2) can be shown by tilting the Doob operator with parameter $s_0$,
\begin{align}
L_{s_0,s}^D[\alpha(x)] &= e^{-\theta({s_0})} l_{s_0}(x) L_{s_0}[e^{-{s} g(x)} \left(l_{s_0}(x)\right)^{-1}\alpha(x)]  = e^{-\theta({s_0})} l_{s_0}(x) L_{s_0+s}[\left(l_{s_0}(x)\right)^{-1}\alpha(x)]\, ,
\end{align}
whose right eigenfunction associated with the largest eigenvalue $e^{\theta^D(s)}$, is $l_{s_0}(x) r_{s_0+s}(x)$, since
\begin{equation}
L_{s_0,s}^D[l_{s_0}(x) r_{s_0+s}(x)]  = e^{-\theta({s_0})} l_{s_0}(x) L_{s_0+s}[r_{s_0+s}(x)] = e^{\theta(s_0+s)-\theta(s_0)} l_{s_0}(x) r_{s_0+s}(x).
\end{equation}
The SCGF of the Doob operator is thus $\theta^D(s) = \theta(s_0+s)-\theta(s_0)$. 
This means that the stationary dynamics of the Doob operator has the same statistics of $A$, given by the derivatives of $\theta^D(s)$ at $s = 0$, which are equal to the derivatives of the SCGF $\theta(s)$ at $s_0$. By Legendre transforming $\theta^D(s)$ we get $I^D(a)=I(a)+\theta(s_0)+s_0 a$, which is the rate function of the tilted distribution, $P_{s_0}(a)\sim e^{-N I^D(a)}$. The atypical fluctuations of the natural dynamics given by Eq.\! (\ref{L}) associated with some $s_0\neq 0$ thus become typical in the Doob-transformed dynamics.

\subsection{Doob effective map}

The Doob operator (\ref{LD}) has a stationary state that naturally yields the statistics for $A$ that correspond to rare fluctuations of the original dynamics for some $s_0\neq 0$. 
For long times, points of the typical trajectories generated from the original dynamics are distributed according to a density that approaches the invariant measure of $f$, which we denote as  $\rho(x)$, whereas points of the rare trajectories approach $\rho^D_{s_0}(y) = l_{s_0}(y) r_{s_0}(y)$.  But how the latter  distribution can be generated from the typical trajectories of an effective chaotic map (which must be necessarily different from $f$) is not clear at this point. Specifically, we would like to find a Doob effective map $f_{s_0}^D$, whose evolution $y_{n+1} = f^D_{s_0}(y_n)$ yields an invariant measure that supports that Doob-operator statistics $\rho^D_{s_0}(y)$. 

To address this question we consider the problem of how to transform values $x$ taken by a random variable $X$ distributed according to the probability density $\rho(x)$ into values $y$ taken by a random variable $Y$ distributed according to a probability density $\rho^D_{s_0}(y)$. Specifically we will look for an invertible function $\gamma_{s_0}$ that carries out this mapping: $y = \gamma_{s_0}(x)$. We assume that $\rho(x)$ and $\rho^D_{s_0}(y)$ are integrable (for example, they are continuous functions or have countably many discontinuity points, so the Riemann integral exists) and strictly positive ($\rho(x)>0$ for all $x\in I$, $\rho^D_{s_0}(y) >0$ for all $y\in \gamma_{s_0}(I)$), so that their cumulative distributions $F(x) = \int_{-\infty}^x \rho(u) du$ and $F^D_{s_0}(y) = \int_{-\infty}^y \rho^D_{s_0}(u) du$ are continuous and increasing (hence invertible) functions. It is clear that
\begin{equation}
F^D_{s_0}(y) = P(Y \leq y) = P(\gamma_{s_0}(X) \leq y) = P(X \leq \gamma_{s_0}^{-1}(y)) = F(\gamma_{s_0}^{-1}(y)) = F(x).
\label{gamma}
\end{equation}
Solving for $y$, the required transformation is found to be $y= \gamma_{s_0}(x) = (F^D_{s_0})^{-1}(F(x))$. Applying this transformation it is straightforward to find the Doob effective map taking into account that $y_{n+1}=f^D_{s_0}(y_{n})=f^D_{s_0}(\gamma_{s_0}(x_{n}))$ and that $y_{n+1}=\gamma_{s_0}(x_{n+1})=\gamma_{s_0}(f(x_{n}))$. From these equations we obtain $f^D_{s_0}(\gamma_{s_0}(x_{n}))=\gamma_{s_0}(f(x_{n}))$, i.e. $f^D_{s_0} \circ \gamma_{s_0}= \gamma_{s_0} \circ f $, so that the Doob effective map, which is topologically conjugate to $f$, takes the form
\begin{equation}
f^D_{s_0} = \gamma_{s_0} \circ f \circ \gamma_{s_0}^{-1},
\label{fD}
\end{equation}
or equivalently $f = \gamma_{s_0}^{-1} \circ f^D_{s_0} \circ \gamma_{s_0}$, with $\gamma_{s_0}$ defined as right below Eq.~(\ref{gamma}). Indeed, mathematically speaking the conjugacy is smoother than that provided by a homeomorphism, as $\gamma_{s_0}$ is differentiable. This amounts to the dynamics given by $f$ plus a change of coordinates $y= \gamma_{s_0}(x)$, such that $y_{n+1} = f^D_{s_0}(y_n) = \gamma_{s_0}(f(\gamma_{s_0}^{-1}(y_n)))$:

\begin{center}
\begin{tikzcd}
x_n \arrow[r, "f"]
& x_{n+1} \arrow[d, "\gamma_{s_0}" ] \\
y_n \arrow[r, "f^D_{s_0}"]\arrow[u, "\gamma_{s_0}^{-1}"]
&  y_{n+1}
\end{tikzcd}
\end{center}

Now, if we generate a long trajectory from an initial value of $x_0$ with the ergodic map $f$, it is clear that for long times the points that make up the trajectory will be distributed according to the invariant measure of $f$, $\rho(x)$. On the other hand, if we transform each of these values $x_0, x_1, x_2, \ldots$ with the function $\gamma_{s_0} = (F^D_{s_0})^{-1} \circ F$, the resulting sequence of states $y_0 = \gamma_{s_0}(x_0), y_1 = \gamma_{s_0}(x_1), y_2 = \gamma_{s_0}(x_2), \ldots$, will be distributed following $\rho^D_{s_0}(y)$. 
The rare fluctuations of $f$, corresponding to a given $s_0 \neq 0$, become typical in $f^D_{s_0}$.

\section{Numerical method to solve the eigenvalue problem}

In order to obtain the Doob effective map of the examples discussed in the main text, we need the largest eigenvalue of the tilted operator, as well as its left and right eigenfunctions. Hence we have numerically solved the eigenvalue problem for the tilted operator given by \eqref{Lsrs} and \eqref{Lsls}. To that end we have extended to general observables a numerical technique developed in Ref.~\cite{feigenbaum1989} for the case of the finite-time Lyapunov exponent, $g(x)=\ln |f'(x)|$, in which the eigenvalue problem \eqref{Lsrs} is iteratively solved using Eq.~\eqref{Ls}, through the following recurrence relation
\begin{equation}
\lambda_r (s)  r_s^{(i+1)}(x)=\sum_{z\in f^{-1}(x)}  \frac{\displaystyle e^{-s g(z)}r_s^{(i)}(z)}{|f'(z)|}\, ,
\label{itpr}
\end{equation}
with $i=0,1,...$. Since the limit of the iterative procedure is unique, we start with an initial function $r_s^{(0)}(x)=1$. As described in  \cite{feigenbaum1989}, at each step $r_s^{(i)}(x)$ is normalized, $\int_I r_s^{(i)}(x) dx=1$, so that $r_s(x)$ is approached by the sequence of functions $r_s^{(i)}(x)$, which evolves as
\begin{equation}
r_s^{(i+1)}(x)=\dfrac{\sum_{z\in f^{-1}(x)}  \frac{\displaystyle e^{-s g(z)}r_s^{(i)}(z)}{|f'(z)|}}{\lambda_r^{(i)}}\, ,
\label{itpr2}
\end{equation}
with
\begin{equation}
\lambda_{r}^{(i)}=\int_I dx\sum_{z\in f^{-1}(x)}  \frac{\displaystyle e^{-s g(z)}r_s^{(i)}(z)}{|f'(z)|}\, .
\label{itpr3}
\end{equation}
The sequences \eqref{itpr2} and \eqref{itpr3} converge to the right eigenfunction $r_s(x)$ and to the largest eigenvalue $\lambda(s) = e^{\theta(s)}$, respectively, as $i$ increases.

We now apply the same strategy to obtain the left eigenfunction [eigenvalue problem \eqref{Lsls}], with the normalization at each step satisfying $\int_I r_s^{(i)}(x)l_s^{(i)}(x)dx=1$, so that
\begin{equation}
l_s^{(i+1)}(x)=\dfrac{e^{-s g(x)} l_s^{(i)}(f(x))}{\lambda_l^{(i)}}\, ,
\label{itpr4}
\end{equation}
with
\begin{equation}
\lambda_l^{(i)}=\int_I dx~e^{-s g(x)} l_s^{(i)}(f(x)) r_s^{(i+1)}(x)\, .
\label{itpr5}
\end{equation}
We start as well with $l_s^{(0)}(x)=1$, and, in this case, the sequences \eqref{itpr4} and \eqref{itpr5} converge to the left eigenfunction $l_s(x)$ and to the largest eigenvalue $\lambda(s)$, respectively, as $i$ increases.

\begin{figure}
\includegraphics[width=1.01\linewidth]{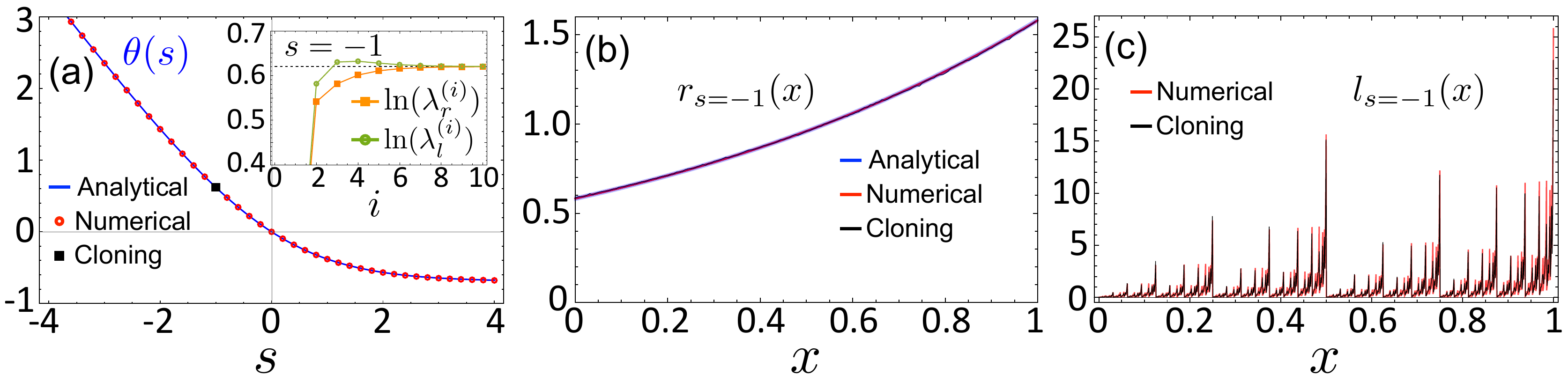}
\caption{
{\bf Comparison between analytical results and numerical results obtained with the cloning algorithm and with the numerical method, given by \eqref{itpr2}-\eqref{itpr5}, used in the main text.} We consider here the fluctuations of the time-averaged position $A=N^{-1}\sum_{n=0}^{N-1} x_n$ for the doubling map $x_{n+1}=2x_n\!\!\!\mod 1$. (a) SCGF: The blue solid line is the analytical result while the red dots have been obtained with the numerical method. The black square is the value obtained with the cloning algorithm for $s=-1$. The inset shows the convergence to the analytical result (black dashed line) of the natural logarithm of the largest eigenvalue associated with the right and left eigenfunctions using the numerical method for $s=-1$.  (b) Analytical result (blue solid line) for the right eigenfunction, together with the results based on the cloning algorithm (red solid line) and the numerical method (black solid line), all for $s=-1$. A perfect overlap is observed. (c) Left eigenfunction for $s=-1$. In the absence of an explicit analytical expression, results with the cloning algorithm (red solid line) and the numerical method (black solid line) are displayed.}
\label{fig1SM}
\end{figure}

This is how the SCGF and the eigenfunctions have been calculated in all the examples of the main text. In order to reach values of $i$ large enough to ensure a clear convergence, we have discretized the spatial coordinate $x$ in equations \eqref{itpr2}-\eqref{itpr5} by dividing the interval $[0,1]$ into $3\times 10^5$ subintervals of equal length. The largest iteration considered $i_{max}$ has been adapted to each particular case: thus, in the tent map example, displayed in Fig.~\ref{fig1} of the main text, we have taken $i_{max}=12$, while in the second example, displayed in Fig.~\ref{fig2} of the main text, we have taken $i_{max}=50$ to ensure a good convergence. In the third example, presented in Fig.~\ref{fig3} of the main text, to construct the Doob effective map for $s=-3$ we have used $i_{max}=10$, while to obtain it at the critical point, $s=-2$, we have taken $i_{max}=100$, since, due to the non-analyticity, the convergence is very slow at that point.

To further illustrate this method, we compare the convergence of the numerical results with an analytical example that has been recently solved \cite{smith22}, namely, the large deviations of the time-averaged position, $g(x)=x$, in the one-dimensional doubling map, $f(x)=2x\!\!\mod 1$, where $x\in[0,1]$ and $z\!\!\mod 1$ is the fractional part of $z$. This chaotic map has an invariant measure $\rho(x)=1$, a SCGF $\theta(s)=\ln[(1+e^{-s})/2]$ and a right eigenfunction equal to $r_s(x)=-s e^{-s x}/(e^{-s}-1)$, as shown in \cite{smith22}.

By applying the numerical method above \eqref{itpr2}-\eqref{itpr5}, we obtain the results displayed in Fig.~\ref{fig1SM}(a), where we observe perfect agreement of the SCGF with the analytical results. The inset to Fig.~\ref{fig1SM}(a) displays the convergence of the natural logarithm of the largest eigenvalue associated with the right and left eigenfunction, i.e., $\ln[\lambda_r^{(i)}]$ and $\ln[\lambda_l^{(i)}]$, for $s=-1$. With $i_{max}=10$ iterations there is already a good agreement with the theoretical value corresponding to the dashed black line.

In addition, we have applied the cloning algorithm developed in \cite{giardina06a,tailleur07} to further check the validity of our results for $s=-1$, specially to corroborate the convergence of the left eigenfunction, for which there is no explicit analytical expression. As we can observe, the SCGF [see black square in Fig.~\ref{fig1SM}(a)], the right eigenfunction [displayed in Fig.~\ref{fig1SM}(b)] and the left eigenfunction [shown in Fig.~\ref{fig1SM}(c)] are all in excellent agreement with the numerical and also the analytical results (when available). This further supports the validity of the numerical method here presented and used in all the examples of the main text.

It is worth noting that computing the left eigenfunction with the cloning algorithm is non-trivial. Indeed, we have to first measure the histogram of the midtime statistics, calculated for the intermediate times of the trajectories sustaining the prescribed rare fluctuation. This is achieved by tracing backwards those trajectories of the clones that survive until the end, as explained in Ref.~\cite{perez-espigares19a}. By doing so, we obtain $p^{mid}_s(x)=l_s(x)r_s(x)$, which corresponds to the Doob stationary state. Then, in order to obtain the left eigenfunction, we need to divide $p^{mid}_s(x)$ by the statistics at the endtime, which yields $p^{end}_s(x)=r_s(x)$. Further details about midtime and endtime statistics can be found in \cite{garrahan09a,nemoto16a,perez-espigares19a,derrida19a}. The parameters of the cloning algorithm in this case are $N_{\text{clones}}=2\times 10^4$ (corresponding to the number of trajectories) starting with a uniform random initial condition. The duration of the trajectories is $N=1000$, and they are averaged over $200$ realizations. A noise of amplitude $\epsilon=10^{-16}$ is added to the dynamics at each time-step (as $\sqrt{\epsilon}\eta_n$, with $\eta_n$ being a white noise of unit variance), since the maps are deterministic and the clones would perform a poor sampling otherwise, see \cite{giardina06a,tailleur07}.

\begin{figure}
\includegraphics[width=1.01\linewidth]{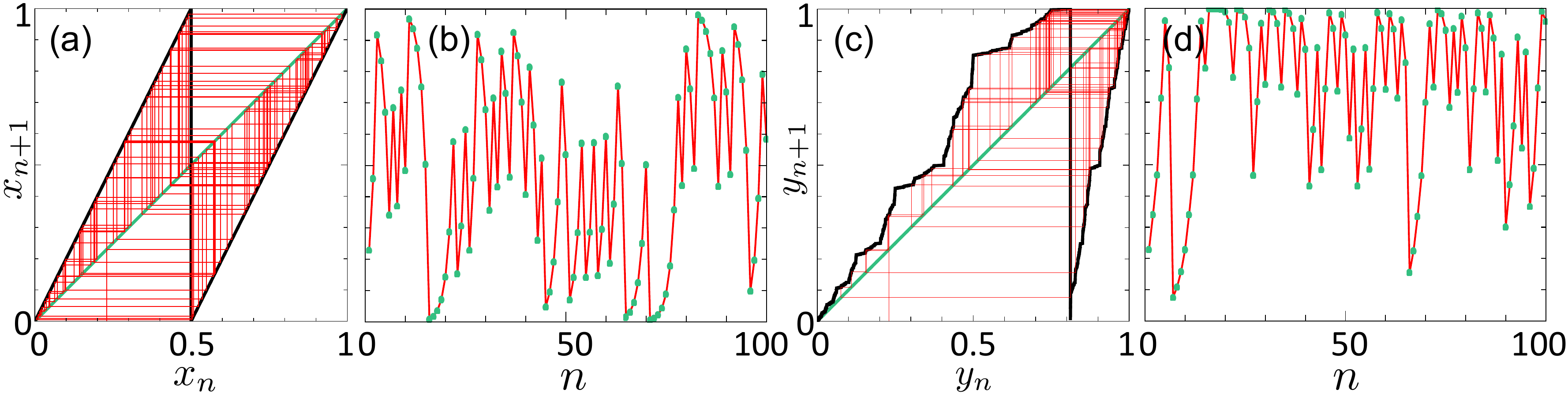}
\caption{
{\bf Original and Doob effective map, together with the typical and atypical trajectories for the doubling map.} Fluctuations of the time-averaged position, $A=N^{-1}\sum_{n=0}^{N-1} x_n$, of the doubling map $x_{n+1}=2x_n\!\!\!\mod 1$. (a) Cobweb plot for $N=100$ iterations of the doubling map. (b) Trajectory illustrated in (a). (c) Cobweb plot for $N=100$ iterations of the Doob effective map for $s_0=-1$. (d) Trajectory illustrated in (c).}
\label{fig2SM}
\end{figure}

Finally, we also show the Doob effective map in this case (with $s_0=-1$). It has been computed from Eq.~\eqref{fD} with $\gamma_{s_0}(x) = (F^D_{s_0})^{-1}(F(x))$, where $F(x) = \int_{0}^x \rho(u) du$ and $F^D_{s_0}(y) = \int_{\gamma_{s_0}(0)}^y \rho^D_{s_0}(u) du$ have been numerically integrated. For the doubling map, the invariant measure is uniform $\rho(x)=1$, and $\rho^D_{s_0}(x)=l_{s_0}(x)r_{s_0}(x)$, where $r_{s_0}(x)$ and $l_{s_0}(x)$ are the functions displayed in Fig.~\ref{fig1SM}(b) and Fig.~\ref{fig1SM}(c) respectively. The resulting Doob effective map, $f^D_{s_0=-1}(x)$, is shown in Fig.~\ref{fig2SM}(c) (black solid line), together with its corresponding cobweb plot (red solid line),  the trajectory being further represented in Fig.~\ref{fig2SM}(d) for $N=100$ iterations. We also included the original map, $f(x)=2x\!\!\!\mod 1$, in Fig.~\ref{fig2SM}(a), with its cobweb, and the corresponding trajectory also for $N=100$, see Fig.~\ref{fig2SM}(b). In the typical trajectories of the Doob effective map, the biasing ($s_0=-1$) promotes fluctuations larger than the average of the original dynamics, which is $\langle A \rangle=\int_0^1 x \rho(x)dx=0.5$, the biased average being $\langle A \rangle_{s_0=-1}=-\theta'(s_0)=\int_0^1 x \rho^D_{s_0=-1}(x)dx\approx 0.73$. This is illustrated in the trajectory of Fig.~\ref{fig2SM}(d), generated with $f^D_{s_0=-1}(x)$, which brings a substantial fraction of points above $0.5$. The value of $0.73$ is obtained by averaging over many such trajectories.

As we have pointed out in the main text, solving the eigenvalue problem in order to obtain the Doob effective process is in general a difficult task. In general it is necessary to make use of numerical methods, such as the iterative approach or the cloning algorithm, for this purpose, as explained above. Recall that in this case, an additional requirement to derive the Doob effective map (one which is absent in the case of stochastic dynamics) is the need to find the right eigenfunction and not only the left one. In this sense, it would be very interesting to extend recent numerical schemes that have been put forward in the context of stochastic systems to chaotic maps. These include the cloning algorithm and transition path sampling complemented with trajectory umbrella sampling \cite{nemoto16a,ray18a,klymko18a}, variational tensor networks \cite{banuls19a,causer22a,causer23a}, or machine learning techniques \cite{Oakes_2020,yan22a}. Such techniques might be particularly useful to tackle higher dimensional maps, continuous-time flows as well as to derive the Doob optimal dynamics at finite times.


\end{document}